\documentclass[10pt, conference]{IEEEtran}
\usepackage{algorithmicx}
\usepackage[ruled,vlined,linesnumbered]{algorithm2e}
\usepackage{hhline}
\usepackage{amsmath,mathtools}
\usepackage{amsfonts,amssymb}
\usepackage{mathrsfs}
\usepackage{gensymb} 
\usepackage{caption}
\usepackage{multirow}
\usepackage{graphicx} 
\usepackage{multirow}
\usepackage{enumitem,color}
\usepackage{algpseudocode}
\begin{document}
%
\title{\huge Learn to Augment Network Simulators Towards Digital Network Twins}

\author{\IEEEauthorblockN{Yuru Zhang, Ming Zhao, Qiang Liu \vspace{-0.16in}}\\
\IEEEauthorblockA{
University of Nebraska-Lincoln\\
qiang.liu@unl.edu}\vspace{-0.3in}
\and
\IEEEauthorblockN{Nakjung Choi \vspace{-0.16in}}\\
\IEEEauthorblockA{
Nokia Bell Labs\\
nakjung.choi@nokia-bell-labs.com}\vspace{-0.3in}
}

\maketitle

\begin{abstract}
Digital network twin (DNT) is a promising paradigm to replicate real-world cellular networks toward continual assessment, proactive management, and what-if analysis.
Existing discussions have been focusing on using only deep learning techniques to build DNTs, which raises widespread concerns regarding their generalization, explainability, and transparency. 
In this paper, we explore an alternative approach to augment network simulators with context-aware neural agents. 
The main challenge lies in the non-trivial simulation-to-reality (sim-to-real) discrepancy between offline simulators and real-world networks.
To solve the challenge, we propose a new learn-to-bridge algorithm to cost-efficiently bridge the sim-to-real discrepancy in two alternative stages.
In the first stage, we select states to query performances in real-world networks by using newly-designed cost-aware Bayesian optimization.
In the second stage, we train the neural agent to learn the state context and bridge the probabilistic discrepancy based on Bayesian neural networks (BNN).
In addition, we build a small-scale end-to-end network testbed based on OpenAirInterface RAN and Core with USRP B210 and a smartphone, and replicate the network in NS-3.
The evaluation results show that, our proposed solution substantially outperforms existing methods, with more than 92\% reduction in the sim-to-real discrepancy.
\end{abstract}

\begin{IEEEkeywords}
End-to-End Network, Digital Network Twin, simulation-to-reality Discrepancy, 
\end{IEEEkeywords}

\section{Introduction}
\label{sec:introduction}

To support emerging applications and diverse scenarios, e.g., extended reality and UAVs, mobile networks are evolving towards densification, distributed, and disaggregation.
The ever-increasingly mobile networks pose non-trivial challenges in network management~\cite{shi2021adapting}, such as resource allocation, admission control, and network slicing, in terms of performance estimation, state understanding, and risk evaluation.
Network simulators (e.g., NS-3 and OMNET++) have been widely adopted to analyze, evaluate, and test network management policies, including both model-based and model-free, such as deep learning (DL) and deep reinforcement learning (DRL)~\cite{zhang2020onrl}.
In common practice, the simulated network will be created to match the real-world network, including architecture, topology, and parameters, such as user mobility and radio channel propagation. 
To maintain the feasible computation complexity of network simulations, a variety of abstraction mechanisms (e.g., SNR-to-BLER mapping) are developed and used in network simulators.
However, the introduced abstraction mechanisms may fail to accurately represent the actual behaviors in real-world networks~\cite{liu2022atlas}.
Besides, network simulators cannot incorporate the nature dynamics of real-world networks, where some of them may be unknown and unobservable, such as instantaneous channel variations and thermal impacts on RF devices.
As a result, recent works have been increasingly revealing that the non-trivial simulation-to-reality (sim-to-real) discrepancy~\cite{zhang2020onrl,liu2022atlas, shi2021adapting}, which could substantially compromise the efficacy of simulation results.


Digital network twin (DNT)~\cite{wu2021digital} is a promising paradigm to replicate real-world networks in offline environments, in terms of fine granularity, synchronicity, and high fidelity.
The attribute of high fidelity requires the DNT to have a minimal sim-to-real discrepancy with respect to real-world networks.
With the DNT, network management policies can be offline tested in terms of performance, safety, and robustness.
For example, the network performance of a DL-based policy can be thoroughly evaluated if its management action will violate the service-level agreement (SLA) of user applications~\cite{liu2022atlas}.
The concept of digital twins have been discussed for years, however, there still lacks a concertized and detailed approach to building a digital twin for arbitrary real-world networks.
Deep learning-based approaches have been paid great attention, where one or more deep neural networks (DNNs) will be created and trained to imitate the behaviors of real-world networks~\cite{lin20236g}.
However, the unresolved concerns of DNNs, including generalization, explainability, and transparency raised, which would constrain its deployment in practice.

In this paper, we explore an alternative approach to building digital network twins for arbitrary real-world networks.
The fundamental idea is to augment existing network simulators by reducing the sim-to-real discrepancy with newly-designed context-aware neural agents. 
The rationale is that, network simulators are built based on rigorous domain knowledge, which generally achieves solid generalization and explainability and transparency.
The main challenge lies in the non-trivial and complex sim-to-real discrepancy between offline simulators and real-world networks.
To solve the challenge, we propose a new learn-to-bridge algorithm to cost-efficiently bridge the sim-to-real discrepancy via two alternative stages.
In the first stage, we design a new cost-aware Bayesian optimization to select configurable states and query their performances in real-world networks.
This is based on our observation that, the sim-to-real discrepancy is highly state-dependent and non-uniform.
In the second stage, we create a neural agent with Bayesian neural networks (BNN) and train it to learn the state context and bridge the probabilistic discrepancy.
In addition, we build a small-scale end-to-end network testbed based on OpenAirInterface RAN and Core with USRP B210 and a smartphone, and replicate the network in NS-3.
The evaluation results show that, our proposed solution substantially outperforms existing methods, with more than 90\% reduction in the sim-to-real discrepancy.

\section{System Model}









%
%



We consider a real-world mobile network comprising multiple base stations (BS) in the radio access network (RAN), a core network (CN), and mobile users.
Besides, an offline network simulator (e.g., NS3 and OMNET++) replicates the real-world network in terms of architecture, topology, and parameters.
For example, the simulation parameters of the network simulator, e.g., link bandwidth and operating spectrum, are set to match that of the real-world network.
In this work, we focus on achieving a high-fidelity DNT, which imitates the network performance in the real-world network, such as latency, throughput, and reliability.

Denote ${s}$ as the configurable state of the mobile network, e.g., resource allocation, scheduling strategy, and user traffic.
Given a configurable state, we can obtain the network performance from the network simulator at a negligible cost, in terms of time consumption, computation complexity, and labor efforts. 
In contrast, we consider the cost of querying the real-world network to be non-trivial, which relates to the configuration of distributed network infrastructures and the collection of network performance over time.
Hence, we denote $c({s})$ as the querying cost for the state ${s}$ and the cumulative cost $ C = \sum_{n=1}^{N} c\left(\mathbf{s}_n \right)$ until the $n$th iteration.
Here, we denote $\mathcal{P}_{r}({s})$ and $\mathcal{P}_{s}({s})$ to represent the network performance under the state $s$ obtained from the real-world network and network simulator, respectively.

Note that, each individual network performance includes a set of values, which are collected throughout the collection time period, e.g., 1 hour. 
To represent the discrepancy between two distributions, we define the sim-to-real discrepancy as $\mathcal{D}_{KL}(\mathcal{P}_{r}({s}) \| \mathcal{P}_{s}({s}))$, where $\mathcal{D}_{KL}$ is the operator of KL-divergence.
Using the KL-divergence, we can evaluate how the offline performance distribution is different from real-world performance distribution.
Therefore, we formulate the digital network twin problem as
\begin{align}
    \mathbb{P}_0:  & \quad \min _{A} & \quad \mathcal{D}_{KL}(\mathcal{P}_{r}({s}) \| \mathcal{P}_{s}({s}) + A({s}) ) \\
    & \text { s.t. } & \sum\nolimits_{n=1}^{N} c\left(\mathbf{s}_n \right) \leq C_{\max},\label{constrain}
\end{align}
where $A({s})$ is the probabilistic offset generated by neural agent under the state ${s}$, and $C_{\max}$ is the budget of cumulative querying cost.


\textbf{Challenges.} 
The challenge in solving the aforementioned problem $\mathbb{P}_0$ primarily resides in the designing the neural agent.
To bridge the sim-to-real discrepancy, the neural agent may need massive training dataset to be aware the high-dim state space and generate the performance offset $A({s})$.
On the one hand, the sim-to-real discrepancy is a black-box function, where no closed-form expression to model the network performance obtained in the real-world system $\mathcal{P}_{r}({s})$ and network simulator $\mathcal{P}_{s}({s})$.
Hence, it is difficult to cost-efficiently build the training dataset, where random and grid selection generally lead to low cost-efficiency.
On the other hand, the sim-to-real discrepancy is highly state-dependent, which means the neural agent should generate contextual performance offsets under high-dim states, which is a non-trivial challenge for existing methods, such as linear regression. 


\section{The Proposed Solution}

\begin{figure}[!t]
	\centering
	\includegraphics[width=3.5in]{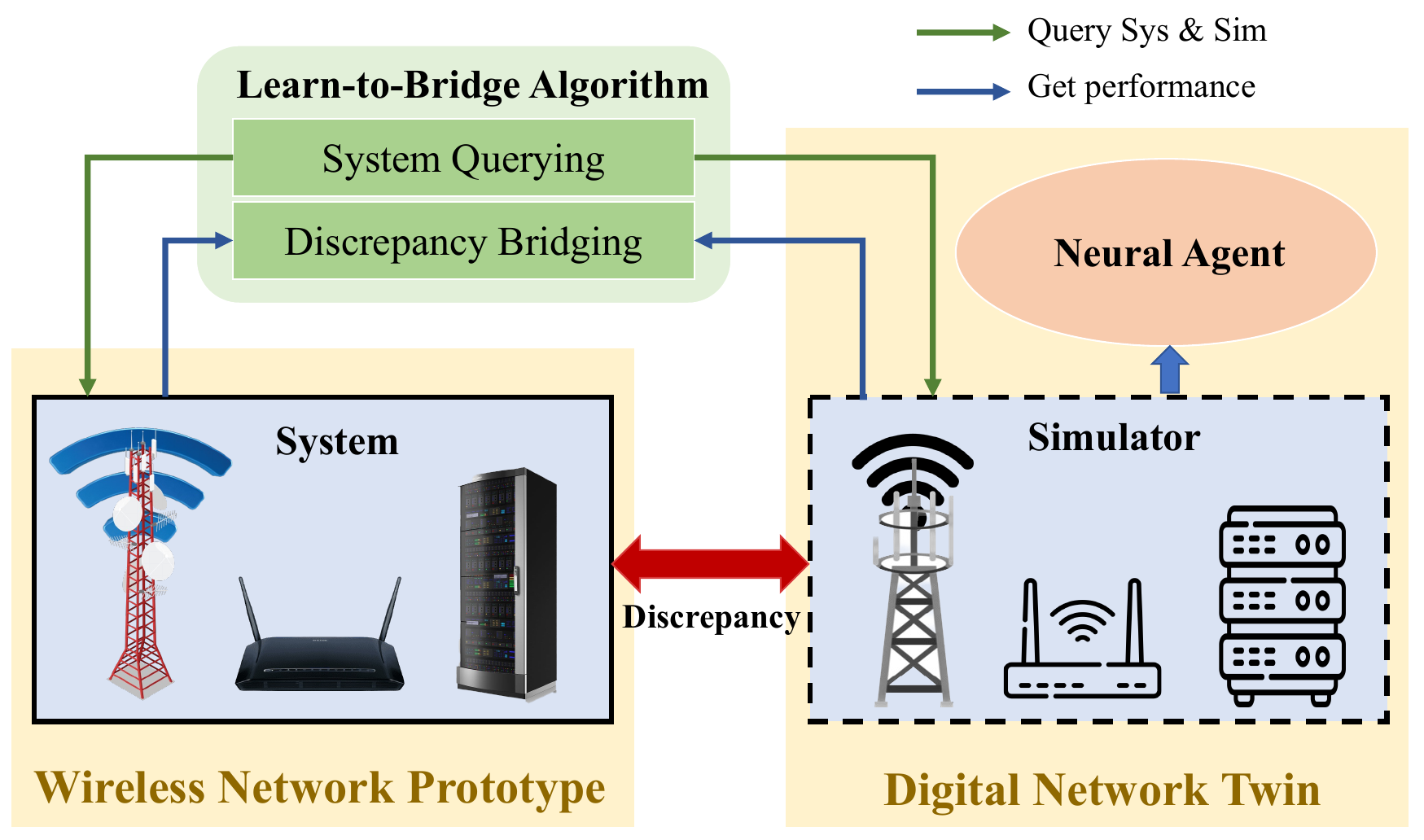}
	\caption{\small The overview of the proposed solution. }
	\label{fig:solution}
\end{figure}

In this section, we describe the proposed digital network twin (DNT) solution (Fig.~\ref{fig:solution}), which augments the network simulator with a newly-designed neural agent.
First, the simulation parameters of the network simulator are configured and calibrated to match that of the real-world network.
Second, we evaluate the sim-to-real discrepancy by sampling states and comparing the corresponding performance collections.
Third, we invoke the proposed learn-to-bridge (L2B) algorithm (see Alg.~\ref{alg:proposed}) to minimize the sim-to-real discrepancy by alternatively selecting states from the whole state space, querying real-world performance collections, and updating the neural agent.
The learn-to-bridge algorithm terminates when the cumulative cost reaches the given threshold.
Finally, we obtain the DNT with the trained neural agent and the network simulator. 

In particular, the proposed learn-to-bridge algorithm is composed of two alternative stages, i.e., the system querying and discrepancy bridging.
In the system querying stage, we select a batch of states and query them in the real-world network by designing a cost-aware Bayesian optimization.
In the discrepancy bridging stage, we update the neural agent to approximate the sim-to-real discrepancy by training a Bayesian neural network (BNN) with latest observations.
By solving these two stages alternatively, the sim-to-real discrepancy would gradually reduced.



\subsection{The Stage of System Querying}
In the system querying stage, we need to select a batch of states and query them in the real-world network.
Here, the objective is to find the maximum sim-to-real discrepancy under the constraint of cumulative costs.
The rationale is that, as long as we found these states with the high sim-to-real discrepancies, the discrepancy bridging stage will train the neural agent to bridge the discrepancy, which maximizes the reduction of sim-to-real discrepancy in this stage.
Thus, the system querying subproblem $\mathbb{P}_1$ can be expressed as 
\begin{align}
    \mathbb{P}_1:  & \quad \max _{\mathcal{S}} & \quad \mathcal{D}_{KL}(\mathcal{P}_{r}({s}) \| \mathcal{P}_{s}({s})) \\
    & \text { s.t. } & \sum\nolimits_{n=1}^{N} c\left(\mathbf{s}_n \right) \leq C_{\max}.
\end{align}
We considering that, conducting the system querying under different states has non-identical costs, such as experiment time, collection efforts, and deployment expenses.
As a result, it is impractical to use grid search based methods to select states from high-dimensional state space (e.g., hundreds of states, if not more) in real-world networks. 
In addition, we observe that the sim-to-real discrepancy is non-uniform and highly state-dependent (See Fig.~\ref{fig:Discrepancy_reduction}).
Hence, the random selection based method generally leads to inefficient searching, because the information from accumulative observations are not fully exploited.
Therefore, we design a new cost-aware Bayesian optimization method to select states.



\textbf{Cost-Aware Bayesian Optimization.}
Bayesian optimization \cite{shahriari2015taking, snoek2012practical} is the state-of-the-art framework to deal with black-box problems, with practically proven sample efficiency and performance. 
It generally consists of a surrogate model to approximate the black-box function and an acquisition function to determine the next action to query.
In each iteration, the surrogate model will be updated with the latest observations for better approximation accuracy, which improves the effectiveness of the action selection by the acquisition function.
Conventional Bayesian optimization mostly focused on unconstrained black-box problems, which fail to tackle the non-identical cost of different states in the problem $\mathbb{P}_0$.
To this end, our proposed cost-aware Bayesian optimization uses Gaussian process (GP) as the surrogate model to sample-efficient approximate the black-box function, and an improved Expected Improvement (EI) to be the acquisition function to cost-efficiently select states.

Expected Improvement (EI) is a widely used acquisition function in a variety of scenarios. 
For the sake of simplicity, we denote $f(\mathbf{s}) = \mathcal{D}_{KL}(
\mathcal{P}_{r}({s}) \| \mathcal{P}_{s}({s}))$ as the black-box function in the problem $\mathbb{P}_0$.
Sample from $f(\mathbf{s})$ to get $\mathbf{s}_t=\operatorname{argmax}_{\mathbf{s}} \mathrm{EI}\left(\mathbf{s} \mid \mathcal{D}_{1:t-1}\right)$, $t\in[1,T]$, where $\mathcal{D}_{1: t-1} = \left\{ \left(\mathbf{s}_1, f\left(\mathbf{s}_1\right)\right), \ldots,\left(\mathbf{s}_{t-1}, f\left(\mathbf{s}_{t-1}\right)\right)\right\}$ is t-1 samples sampled from $f\left(\mathbf{s}\right)$. Here, EI is defined as
\begin{equation}
\mathrm{EI}(\mathbf{s})=\mathbb{E} [\max \left(f\left(\mathbf{s}^{+}\right)-f(\mathbf{s}), 0\right)],
\end{equation}
where $\mathbf{s}^{+}=\operatorname{argmax}_{\mathbf{s}_t \in \mathbf{s}_{1:t}} f\left(\mathbf{s}_t\right)$, $t\in[1,T]$ and $f\left(\mathbf{s}^{+}\right)$ represents the best sample value at present.
At each iteration, the acquisition function selects the point with the highest value.
To consider the cost of states during selection, we design a new cost-aware expected improvement (cEI) as
\begin{equation}
\operatorname{cEI}(\mathbf{s}) = \frac{\mathrm{EI}(\mathbf{s})}{c(\mathbf{s})^{\alpha}},
\label{eq:cEI}
\end{equation}
which evaluates the ratio between the EI value and the cost.
This would encourage the acquisition function to select the states with high EI values and low costs. 
Besides, we introduce a parameter $\alpha \in [0, 1]$ to control the impact of costs to balance the exploration and exploitation during the searching of Bayesian optimization. 
In particular, we use the decrease of sim-to-real discrepancy as the indicator to adjust the parameter $\alpha$.
When the sim-to-real discrepancy decreases quickly in the last few iterations, we adjust the parameter $\alpha$ to be large (e.g., approaching the maximum value 1), where the cEI would become more conservative to exploit existing observations.
Otherwise, we adjust the parameter to be small (e.g., approaching the minimum value 0), where the cEI would trend to explore new states.


We use the sample-efficient Gaussian Process (GP) as the \textit{surrogate model} to approximate the black-box function $f(\mathbf{s})$. GP is an extension of the multivariate Gaussian distribution to infinite dimensions and it can be expressed as $f(\mathbf{s}) \sim \mathcal{G} \mathcal{P}\left(\mu(\mathbf{s}), K\left(\mathbf{s}, \mathbf{s}^{\prime}\right)\right)$. It represents the distribution of functions $f(\mathbf{s})$ and consists of both mean function $\mu(\mathbf{s})$ and covariance function $K\left(\mathbf{s}, \mathbf{s}^{\prime}\right)$. The covariance of GP can be expressed using a kernel function. In this case, we utilize the widely used Radial Basis Function (RBF) as the kernel function, which is defined as $K\left(\mathbf{s}, \mathbf{s}^{\prime}\right)=\sigma^2 \exp \left(-\frac{\left\|\mathbf{s}-\mathbf{s}^{\prime}\right\|^2}{2 l^2}\right)$. Here, the hyper-parameters of the Gaussian kernel, the variance $\sigma$ and the length $l$, determine the average distance from the function mean and the extent of influence on neighboring points, respectively, and $\left\|\mathbf{s}-\mathbf{s}^{\prime}\right\|^2$ denotes the distance between two distinct points in the continuous domain of the GP. After t collections, we can get $\mathcal{D}_{1:t} = \left\{ \left(\mathbf{s}_1, f\left(\mathbf{s}_1\right)\right), \ldots,\left(\mathbf{s}_{t}, f\left(\mathbf{s}_{t}\right)\right)\right\}$. The posterior distribution is expressed as $P\left(f(\mathbf{s}) \mid \mathcal{D}_{1:t}\right) \propto P\left(\mathcal{D}_{1:t} \mid f(\mathbf{s})\right) P(f(\mathbf{s})) $. Since any point $f\left(\mathbf{s}_{t+1}\right)$ on GP and previous observation data follow a joint Gaussian distribution, we can further obtain the predicted distribution as
\begin{equation}
P\left(f\left(\mathbf{s}_{t+1}\right) \mid \mathcal{D}_{1: t}, \mathbf{s}_{t+1}\right)=\mathcal{N}\left(\mu_t\left(\mathbf{s}_{t+1}\right), \sigma_t^2\left(\mathbf{s}_{t+1}\right)\right).
\end{equation}


\subsection{The Stage of Discrepancy Bridging}
In the discrepancy bridging stage, we aim to update the neural agent to minimize the sim-to-real discrepancy, according to the accumulative observation of performance collections.
Note that, the sim-to-real discrepancy is calculated based on the performance collection in both simulator and real-world system, which are a set of values, rather than a fixed value.
For example, we can obtain the collection of end-to-end latency of user application (e.g., hundreds of values) under each state state.
Hence, the neural agent needs to bridge the probabilistic discrepancy under all states.
Although Gaussian process (GP) is a promising method to approximate probabilistic distributions, it can only generate Gaussian distribution, while the sim-to-real discrepancy is barely Gaussian under all states.
To this end, we adopt Bayesian Neural Network (BNN) to approximate the sim-to-real discrepancy function $\mathcal{D}_{KL}(\mathcal{P}_{r}({s}) \| \mathcal{P}_{s}({s}))$.





\textbf{Bayesian Neural Networks.}
By introducing the prior probability for weight $\mathbf{w}$ of BNN, the objective of training the BNN model transforms into finding the Maximum a Posteriori (MAP), the posterior distribution of weight $\mathbf{w}\sim P(\mathbf{w}\mid \mathcal{D})$, where $\mathcal{D}$ refers to the observed dataset. According to Bayesian theory, the calculation of the posterior probability distribution is:
\begin{equation}
P(\mathbf{w} \mid \mathcal{D})=\frac{P(\mathbf{w}, \mathcal{D})}{P(\mathcal{D})}=\frac{P(\mathcal{D} \mid \mathbf{w}) P(\mathbf{w})}{P(\mathcal{D})}.
\end{equation}
Here, we need to obtain both the prior probability $P(\mathbf{w})$ and the likelihood probability $P(\mathcal{D} \mid \mathbf{w})$. However, calculating the posterior probability is intractable. Consequently, we approximate the posterior probability by defining a probability distribution $q_\theta(\mathbf{w} \mid \mathcal{D})$, which serves as a substitute for $p(\mathbf{w} \mid \mathcal{D})$. Thus, our goal is to minimize the distance between $q_\theta(\mathbf{w} \mid \mathcal{D})$ and $p(\mathbf{w} \mid \mathcal{D})$, and we employ KL-divergence to quantify this distance:
\begin{equation}
\begin{aligned}
\theta^{\text {opt }} &=\arg \min _\theta \mathcal{D}_{KL}(q_\theta(\mathbf{w} \mid \mathcal{D}) \| p(\mathbf{w} \mid \mathcal{D})) \\
& = \arg \min _\theta \int q_\theta(\mathbf{w} \mid \mathcal{D}) \log \frac{q_\theta(\mathbf{w} \mid \mathcal{D})}{p(\mathbf{w} \mid \mathcal{D})} dw.
\end{aligned}
\end{equation}

\subsection{The Learn-to-Bridge Algorithm}

\begin{algorithm}[!t]
    \caption{The Learn-to-Bridge (L2B) Algorithm}\label{alg:proposed}
    \KwIn{$T$, $C_{\max}$}
    
    \While{$True$}{
        
        $/**\;System\; Querying\; Stage**/$\;
        \For{$t = 0,1,...,T$}{
            Sample states from $\mathcal{S}$ except previously observed states\;
            
            Estimated their costs $c\left(\mathbf{s} \right)$\;

            Calculate their cEI values based on Eq.~\ref{eq:cEI}\;
            
            Select the optimal state $\mathbf{s}$\;

            Query both simulator and real-world system\;
            
            Calculate sim-to-real discrepancy $\mathcal{D}_{KL}(\mathcal{P}_{r}(\mathbf{s}_t) \| \mathcal{P}_{s}(\mathbf{s}_t))$\;

            Update GPR with all cumulative observations\;
            
            Store $<\mathbf{s}_t, \mathcal{D}_{KL}(\mathcal{P}_{r}(\mathbf{s}_t) \| \mathcal{P}_{s}(\mathbf{s}_t))>$\;
        }
        
        Calculate cumulative cost $C=\sum_{t=1}^{T} c\left(\mathbf{s}_t \right)$\;
        
        $/**\;Discrepancy\; Bridging\; Stage**/$\;
        
        Train BNN with $<\mathbf{s}_t, \mathcal{D}_{KL}(\mathcal{P}_{r}(\mathbf{s}_t) \| \mathcal{P}_{s}(\mathbf{s}_t))>$\;


        
        \If{Cumulative cost exceeds the threshold}{
            break\;
        }
    }
    
\end{algorithm}

With the aforementioned analysis, we summarize the learn-to-bridge (L2B) algorithm in Alg.~\ref{alg:proposed}.

In the system querying stage, we iteratively select and query state. 
First, we sample thousands of states from the state space, and estimate their costs $c\left(\mathbf{s} \right)$. 
Next, we calculate the cEI for all sampled state, and choose the optimal state with highest cEI value.
Then, we query the selected state and calculate the sim-to-real discrepancy $\mathcal{D}_{KL}(\mathcal{P}_{r}(\mathbf{s}_t) \| \mathcal{P}_{s}(\mathbf{s}_t))$, which will be used to update the GPR in the cost-aware Bayesian optimization.
As the batch of states are selected, we train the BNN to approximate the sim-to-real discrepancy.
When the cumulative cost surpluses the given threshold $C_{\max}$, the algorithm terminates.

\section{System Implementation}
In this section, we describe the system implementation, including the end-to-end testbed and the proposed DNT.

\subsection{End-to-End Testbed}
We implement an end-to-end network testbed, including a eNB in RAN, a CN, and a mobile user. 
We implement CN based on OpenAir-CN with the separation of the control plane and data plane (CUPS), wherer network functions (NFs), e.g., HSS, SPGW-U and SPGW-C, are deployed using Docker containers. 
We implement the RAN based on OpenAirInterface (OAI) on an Intel i7 desktop with Ubuntu 18.04 low-latency kernel. 
The RAN host is connected to an USRP B210 serving as the RF front end. 
The eNB operates on frequency band 7 with a 10MHz radio bandwidth. 
An Oneplus 9 Android smartphone is connected to eNB as the UE, with 1 meter UE-to-Antenna distance.
We develop an Android application that continuously sends video frames to edge server co-located with the SPGW-U. 
These video frames are processed by the server using a feature extraction algorithm (i.e., ORB), and the results are subsequently fed back to the UE. 
The performance metric for evaluating this application is the end-to-end latency of frames. 

The \textbf{state} $\mathcal{S}= \left\{ U, D, C, R, M_{U}, M_{D},F \right\} $ includes 1) $ U\in[0,50]$ is the uplink bandwidth allocation; 2) $D\in[0,50]$ is downlink bandwidth allocation; 3) $C\in[0,1.0]$ is CPU ratio; 4) $R\in [0,1.0]$ is RAM ratio; 5) $M_{u}\in[0,20]$ is average uplink MCS; 6) $M_{d}\in[0,28]$ is average downlink MCS; 7) $F\in[1,4]$ is user traffic. 
Note that, the state space is discrete due to the intrinsic implementation of the end-to-end testbed, such as physical resource block (PRB) allocation in the MAC layer and the number of user traffic.
Hence, we exhaustively search the whole state space and generate a database of network performance with nearly 2500 states.

\begin{figure}[!t]
	\centering
	\includegraphics[width=3.4in]{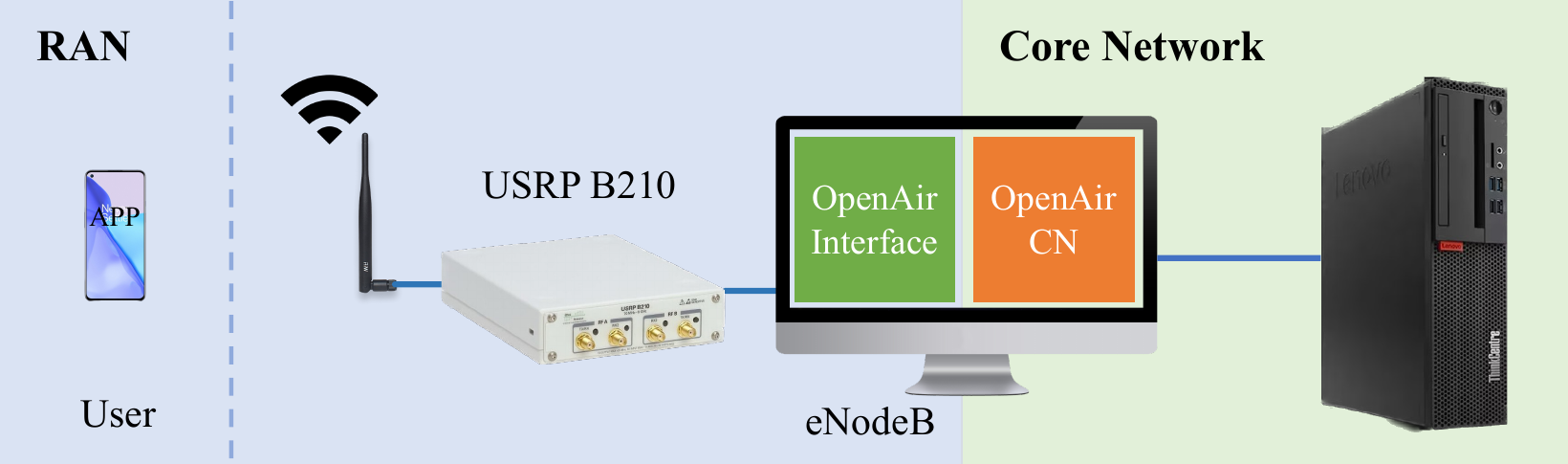}
	\caption{\small The overview of the end-to-end network testbed. }
	\label{fig:DNT-testbed}
\end{figure}

\subsection{Digital Network Twin}

The proposed DNT is mainly composed of an offline network simulator and a neural agent.

\textbf{Simulator.}
We replicate the end-to-end testbed by using Network Simulator 3 (NS-3), where the simulation parameters are matched correspondingly.
In particular, we adopt the \emph{LogDistancePropagationLossModel} as the pathloss model while omitting any fading models. 
For the transport network, a p2p link is created to connect RAN and CN, where the bandwidth and delay are configured based on realistic measurements. 
We also replicate the traffic generation of mobile users and the edge computing processing by developing a FIFO service queue.
Specifically, the transmission sizes match closely with a mean of 28.8kb and a standard deviation of 9.9kb. 
Besides, other simulation parameters are calibrated with that of the end-to-end testbed, including the MAC scheduler algorithm, antenna specifications, frequency band, and the distance between the eNB and smartphones. 
The end-to-end latency of frames are logged and extracted as the output of network simulations.

\textbf{Neural Agent.}
We implement the neural agent with a BNN model with 4-layer fully connected architecture (i.e., 128x256x256x128), in PyTorch.
We use the \textit{Tanh} activation function in the BNN \cite{Goodfellow-et-al-2016}. 
We utilize the \textit{Adadelta} optimizer with the initial learning rate of 0.001, where the learning rate is decayed by using the \textit{StepLR} scheduler with gamma 0.95. 

\textbf{Learn-to-Bridge Algorithm.}
We use cost-aware Bayesian optimization in the first stage, where GP is implemented by using \textit{scikit-learn} with the \textit{GaussianProcessRegressor} module. 
We utilize the \textit{Matern} kernel with $nu = 2.5$, which extends the Radial Basis Function (RBF) kernel and the absolute exponential kernel by introducing the parameter $nu$.
The $nu$ parameter governs the smoothness of the learned function. 


\section{Performance evaluation}









\begin{figure*}[!t] 
\captionsetup{justification=centering}
  \begin{minipage}[t]{0.33\textwidth}
	\centering
	\includegraphics[width=2.5in]{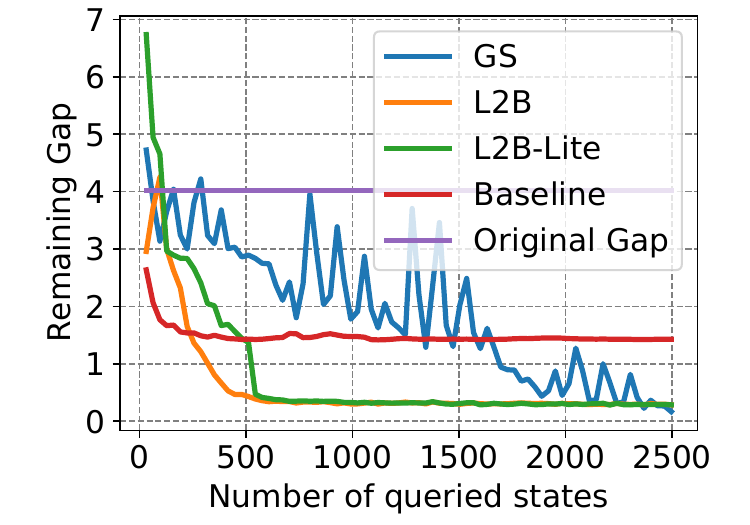}
	\vspace{-0.2in} \caption{\small Performance of discrepancy reduction.}
	\label{fig:gap_reduction}
  \end{minipage}
  \begin{minipage}[t]{0.33\textwidth}
	\centering
	\includegraphics[width=2.5in]{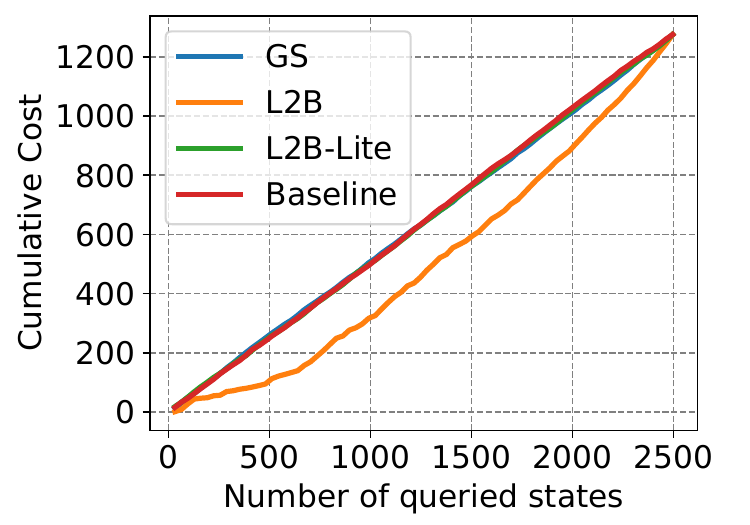}
	\vspace{-0.2in} \caption{\small Performance of cumulative cost.}
	\label{fig:cumulative_cost}
  \end{minipage}
  \begin{minipage}[t]{0.33\textwidth}
	\centering
	\includegraphics[width=2.5in]{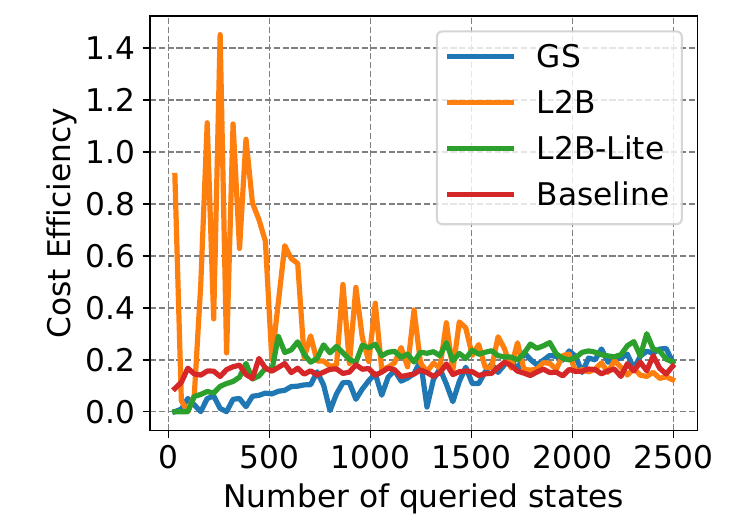}
	\vspace{-0.2in} \caption{\small Performance of cost efficiency.}
	\label{fig:cost_efficiency}
  \end{minipage}
\end{figure*}

\begin{figure*}[!t] 
\captionsetup{justification=centering}
  \begin{minipage}[t]{0.49\textwidth}
	\centering
	\includegraphics[width=2.5in]{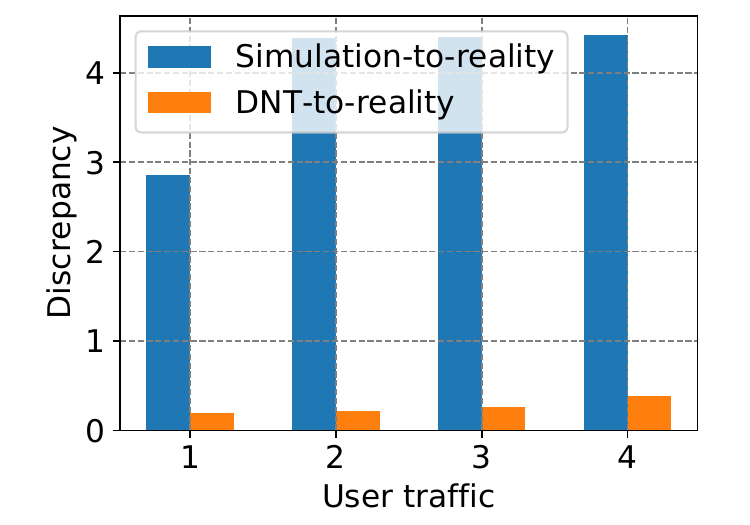}
    \caption{\small Discrepancy reduction under different user traffic.}
	\label{fig:gap_remaining_gap}
  \end{minipage}
  \begin{minipage}[t]{0.49\textwidth}
	\centering
	\includegraphics[width=2.5in]{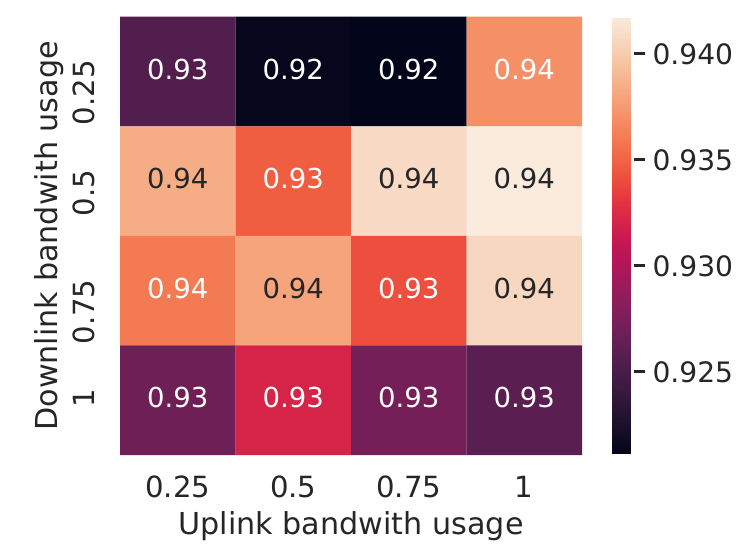}
   \caption{\small Discrepancy reduction under different states.}
	\label{fig:Discrepancy_reduction}
  \end{minipage}
\end{figure*}

In the performance evaluation, we compare our proposed algorithm with the following methods:
\begin{itemize}
    \item \textbf{Baseline}: The \emph{Baseline} randomly selects states to query the real-world network, and use linear regression to bridge the sim-to-real discrepancy.
    \item \textbf{Grid Search (GS)}: The \emph{GS} relies on grid search to select states to query the real-world network, and use Gaussian process to bridge the sim-to-real discrepancy.
    \item \textbf{L2B-Lite}: The \emph{L2B-Lite} is the simplified version of our proposed L2B algorithm, while it is unaware of the cost of different states during the system querying stage. 
\end{itemize}

Fig.~\ref{fig:gap_reduction} shows the performance of different methods for bridging the sim-to-real discrepancy. 
Here, we show the original sim-to-real discrepancy, which is calculated by averaging the sim-to-real discrepancy of all states in the database.
It can be seen that, all methods can gradually reduce the sim-to-real discrepancy as the number of queried states accumulate.
This is because, more queried states would help to continuously improve the accuracy of approximating the discrepancy.
In particular, our proposed learn-to-bridge algorithm achieves the fastest reduction of the discrepancy, where the sim-to-real discrepancy is reduced by nearly 90\% with only 500 queried states.
Note that, the achieved sim-to-real discrepancy between our algorithms and \emph{GS} method are almost the same, when all the states in the database are queried.
This can be attribute to the limited size of the database, where Gaussian process can achieve similar approximation accuracy of the sim-to-real discrepancy with Bayesian neural networks (BNN).
In large-scale operating network, we expect that the Gaussian process could be insufficient to tackle the high-dim state space.

Fig.~\ref{fig:cumulative_cost} shows the accumulation of the querying cost versus the number of queried states.
It can be seen that, our proposed learn-to-bridge algorithm achieves the lowest accumulated querying costs, which is almost 40\% less than all other methods after 500 states are queried.
Note that, all other methods (including \emph{L2B-Lite}) are unaware of the querying cost when they select the states, hence we observe the accumulated cost grows linearly. 
In particular, we show the cost efficiency of all methods in Fig.~\ref{fig:cost_efficiency}, which is defined as the ratio between the reduced sim-to-real discrepancy and the accumulated querying cost.
We can see that, our proposed learn-to-bridge algorithm substantially outperforms all other methods (e.g., more than 10x than the \emph{GS} method) before 1000 queried states, which verifies its cost-efficiency in bridging the sim-to-real discrepancy.
After 1500 queried states, the cost-efficiency of all methods are similar.
This is because, the remaining state space keeps shrinking as more states are queried and thus unavailable to be selected at the system querying stage. 
As a result, the learn-to-bridge algorithm has to select these cost inefficient states from the ever-decreasing state space.


Next, we show the achieved performance of the learn-to-bridge algorithm in terms of reducing the sim-to-real discrepancy under different states.
Fig.~\ref{fig:gap_remaining_gap} shows the discrepancy between sim-to-real and DNT-to-reality under different user traffic. 
It can be seen that, our proposed DNT has a minimal discrepancy with the real-world network across all different user traffic scenarios.
Besides, our proposed DNT reduces the discrepancy between simulation and reality by 93.1\%, 95.1\%, 94.0\%, and 91.2\% for 1, 2, 3, and 4 users, respectively.
Fig.~\ref{fig:Discrepancy_reduction} shows the reduction of the sim-to-real discrepancy under different uplink and downlink bandwidth allocation. 
We observe that, the sim-to-real discrepancy is reduced on average of 93.3\%, where at least 92.1\% reduction can be achieved.
From this result, we also justify that the sim-to-real discrepancy is state-dependent, rather than uniform.
Note that, the sim-to-real discrepancy cannot be completely reduced due to the various abstraction of network simulators and unobservable variabilities in real-world networks.





\section{Related work}
\textbf{Digital Network Twin.} The diagram of digital twin has gained extensive attentions with increasingly research efforts on creation, maintenance, and update.
Almeida \emph{et. al}~\cite{almeida2022machine} proposed a machine learning based propagation loss module for NS-3, which enables accurate prediction of propagation loss in real-world environments and replicates the experimental conditions of a specified testbed. Consequently, it facilitates the creation of a digital twin of the wireless network environment within NS-3, allowing for advanced simulations and analysis. 
Tuli \emph{et. al}~\cite{tuli2021cosco} leveraged the accuracy of predictive digital twin models and simulation capabilities by developing a coupled simulation and container orchestration framework. This paper also create a hybrid simulation-driven decision-making approach to optimize network Quality of Service (QoS) parameters. 
Lai \emph{et. al}~\cite{lai2023deep} introduced an algorithm capable of precisely forecasting the future traffic of the target network. This advancement contributes to the development of the digital twin network, facilitating a greater resemblance between the twin and the ontology.
However, exiting works lack the concertized approach to achieve digital twin in networking area.
In contrast, we focus on designing a detailed approach to achieve DNT by augmenting network simulators with context-aware neural agents.

\textbf{Sim-to-Real Discrepancy.}
The sim-to-real discrepancy has been increasingly revealed in networking domain by recent works. 
Shi \emph{et. al}~\cite{shi2021adapting} proposed to adopt domain adaptation to bridge the discrepancy between the simulator and the system, where a DNN model has been designed to effectively transfer the state knowledge learned from the simulator to the real-world network. 
Liu \emph{et. al}~\cite{liu2022atlas} identified the non-trivial sim-to-real discrepancy in network slicing system, and proposed a three-stage approach by bridging the discrepancy, training offline DL policy, and learning online safe DL policy. 
To tackle the reality gap, Zhang \emph{et. al}~\cite{zhang2020onrl} proposed a hybrid learning algorithm and a learning aggregation mechanism to safely and robustly deploy online Deep Reinforcement Learning (DRL) agents in real-world video telephony. 
Tuli \emph{et. al }~\cite{tuli2022simtune} proposed a framework to bridge the reality gap between simulated and real network QoS, by utilizing a low-fidelity neural network-based proxy model to optimize the parameters of a high-fidelity simulator.
However, these works mostly identified the sim-to-real discrepancy and only focused on designing different techniques to adapt to it. 
In this work, we focus on how to cost-efficiently decrease the sim-to-real discrepancy between the proposed DNT and real-world networks.


\section{Conclusion}
In this paper, we presented a new approach to build digital network twin (DNT) by augmenting network simulators with context-aware neural agents. 
To tackle the non-trivial sim-to-real discrepancy, we proposed a new learn-to-bridge algorithm to cost-efficiently select states in the first stage and update the neural agent with accumulative observations in the second stage.
We built a small-scale end-to-end network testbed based on OAI RAN and Core with USRP B210 and a smartphone, and replicate the network in NS-3.
The evaluation results shown that, our proposed solution substantially outperforms existing methods, with more than 90\% reduction in the sim-to-real discrepancy.

\bibliographystyle{IEEEtran}
\bibliography{ref/reference}

\end{document}